\documentclass[pre,11pt,showpacs,showkeys]{revtex4-1}

\usepackage{amsmath,amssymb}    
\usepackage{graphicx}   
\usepackage{verbatim}   
\usepackage{color}      
\usepackage{subfigure}  
\usepackage{hyperref}   
\usepackage[mathscr]{euscript}
\newcommand{\kB}{k_{\mbox{\tiny \it B}}}        
\newcommand{\Ll}{{\mathcal L}}
\newcommand{\gs}{{\mathscr G}}
\newcommand{\es}{{\mathscr E}}
\newcommand{\tr}{\mbox{\rm tr}}                 
\newcommand{\Pro}{\mbox{\rm Prob}}              
\newcommand{\hLam}{{\widehat{\Lambda}}}
\newcommand{\zs}{{\mathscr Z}}
\newcommand{\tlc}{\tau_{\mbox{\em \tiny LC}}}   
\newcommand{\EM}{{\mathbb E}}
\newcommand{\RM}{{\mathbb R}}
\newcommand{\ZM}{{\mathbb Z}}
\newcommand{\Var}{\mbox{\rm Var}}               
\newcommand{\ls}{{\mathscr L}}
\newcommand{\JM}{{\mathbb J}}
\newcommand{\CM}{{\mathbb C}}
\newcommand{\AM}{{\mathbb A}}
\newcommand{\Hh}{{\mathcal H}}
\newcommand{\hp}{{\widehat{p}}}

\begin{document}

\title{Anankeon theory and viscosity of liquids: a toy model}
\author{Jean V. Bellissard}
\affiliation{Georgia Institute of Technology, School of Mathematics, School of Physics, Atlanta, GA 30332, USA}
\affiliation{Fachbereich Mathematik und Informatik,
Westf\"alische Wilhelms-Universit\"at, M\"unster, Germany}

\email[E-mail: ]{jeanbel@math.gatech.edu}

\thanks{This work was funded partially by the SFB 878, M\"unster. The author is especially indebted to T.~Egami, from Oak Ridge National Laboratory and to J.~Langer from UC Santa Barbara, for having patiently explained their understanding and knowledge of this difficult subject. He thanks the members of the Egami group, in particular Y.~Fan,T.~Iwashita, for sharing their results.}

\date{\today}

\begin{abstract}
A simplistic model, based on the concept of anankeon, is proposed to predict the value of the viscosity of a material in its liquid phase. As a result, within the simplifications and hypothesis made to define the model, it is possible to predict (a) the existence of a difference between strong and fragile liquids, (b) a fast variation of the viscosity near the glass transition temperature for fragile liquids. The model is a mechanical analog of the Drude model for the electric transport in metal.
\end{abstract}

\pacs{61.20.Gy., 64.70.pe.ph.pm, 66.20.Cy}

\keywords{viscosity; fragile, strong liquids; anankeon.}

\maketitle

\section{A Deep Unsolved Problem}

This article is proposing a simplistic model to account for the behavior of the viscosity of liquids as a function of temperature. The best illustration of this behavior is the data from Angell's review paper \cite{Ang95}, that can be found below in Fig.~\ref{visc.fig-ang}. It shows on one graph the behavior of the logarithm of the viscosity $\eta$ as a function of inverse temperature, in order to reveal whether or not an Arrhenius law  is valid $\eta \sim e^{W/\kB T}$. The excitation energy $W$ is represented by the slope of the curve in such coordinate axis. The graph actually shows that the Arrhenius law is valid for a category of materials called {\em ``strong liquids''}. For most other materials, called {\em ``fragile''}, the slope changes as the temperature approach the glass transition.

\begin{figure}[ht]
   \centering
\includegraphics[width=9cm]{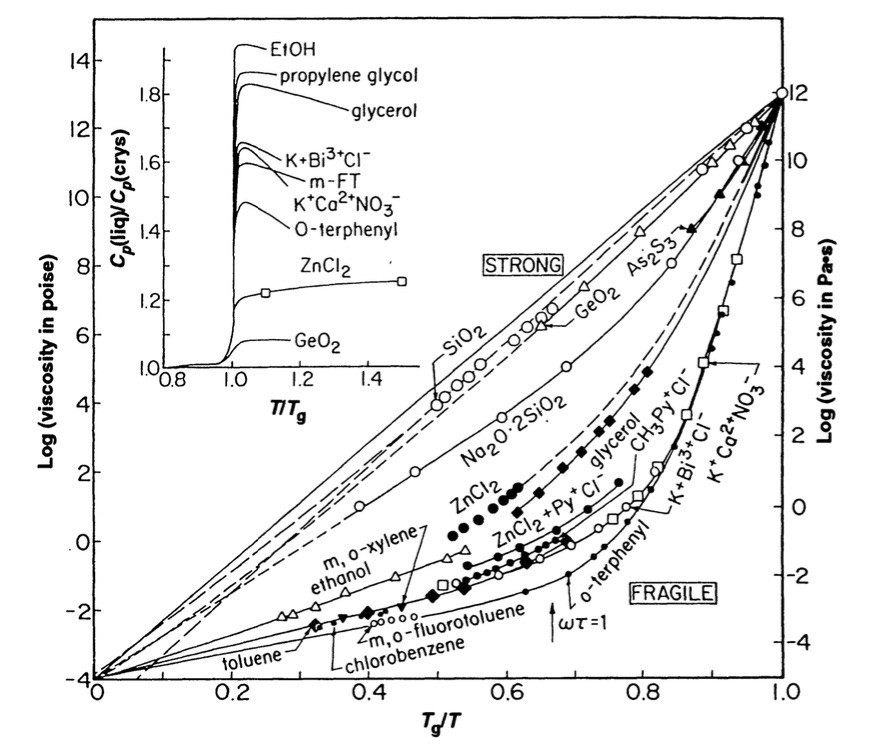}
\caption{\label{visc.fig-ang} Viscosity of the liquid phase of various glassy materials. The logarithm of the viscosity is maps as a function of the normalized inverse temperature, to exhibit Arrhenius law. The normalization is provided by the glass transition temperature $T_g$. (Taken from \cite{Ang95}).}
\end{figure}

The glassy and liquid states of a material share many features. It is understood that the difference between them cannot be account for by using equilibrium Thermodynamics: the glass transition is not a phase transition, it is a dynamical process in which the viscosity becomes so large that the typical time scales involved in the material become macroscopic. The viscosity changes by 15 orders of magnitude over a very small temperature range near the transition. The reason why viscosity changes so rapidly, while the atomic structure does not show signs of this change, has been a mystery for a long time. It has been considered as one of the deepest unsolved problems in science \cite{An95,La07} by the most prominent experts.

 The present model is based upon several decades of experimental observations, numerical simulations and attempts to built a theory liable to explain the origin of viscosity. The author is especially indebted to T.~Egami, from Oak Ridge National Laboratory and to J.~Langer from UC Santa Barbara, for having patiently explained their understanding and knowledge of this difficult subject.

 The work presented here is a small brick added to the knowledge, hoping that it will open the door to a more accurate theory liable to explain this mystery. However, several evidences point toward this model providing some universal results: (a) it introduces a parameter that permits to distinguish between strong and fragile glasses, (b) it exhibits a {\em ``cross-over temperature''} $T_{co}$, below which the viscosity varies very quickly {\em w.r.t.} the temperature. In addition, while being first motivated by the study of bulk metallic glasses, it seems to be relevant in other areas of Physics and Material Sciences, such as colloids \cite{Ega17}, usual glassy materials in their liquid phase, or even quantum fluids \cite{Hel17}.

\subsection{Anankeons}
 \label{visc.ssect-ananke}

The fundamental concept on which this work is based will be called {\em anankeon}. The name came from a long discussion between the present author and T.~Egami, during which the later formulated his view on what were the most fundamental degrees of freedom in the liquid phase \cite{EB12}. More precisely, in his view, it is the cutting and forming of the atomic bond which matters, an action changing the local topology of atomic connectivity network. To explain quickly what the term anankeon means, it suffices to understand that a liquid, like a solid, is a condensed material. Namely the atoms are jam packed together in a local structure made of well ordered clusters, with not much empty places available around. The local stress felt by atoms leads them to jump or to locally reorder themselves  to find a more comfortable position, trying to minimize their energy. These events are fast. They correspond to crossing a saddle point in the energy landscape, namely a passage through an unstable equilibrium positions \cite{Ega14}. In practice they happen at unpredictable times, so that their dynamic is more efficiently described in terms of a Markov process. Hence the corresponding stress tensor felt by each atom is constantly varying under the {\em stress of circumstances}, a concept that can be translated by the word $\alpha\nu\alpha \gamma\kappa\epsilon\iota\alpha$ (anagkeia) in Greek. Associated with this concept was the Greek goddess {\em Ananke}, expressing the fate or destiny, due to uncontrollable forces, constraints, necessity. Hence the name anankeon associated with these stress induced unpredictable atomic moves.

 \subsection{Local Structure, Local Dynamics}
 \label{visc.ssect-loc}

 If there is only one atomic species in a condensed material, the local clusters are likely to have the shape of a regular icosahedron. Since regular icosahedra cannot tile the $3D$-space, there is either a geometrical frustration or the local cluster deform in the solid phase to lead to a crystal. The most likely symmetry of such a crystal is {\em fcc}. To get a glass,  frustration is necessary. In the case of metallic glasses, frustration is obtained by mixing several species with very different atomic radius in order to prevent the systems to crystallize in a periodic order. However, Egami insisted for a long time that this geometrical description of the atomic distribution, while useful, is insufficient at describing the material. He insisted that an atom is not a hard sphere, but exhibits a bit of elasticity. Consequently, an atom can be seen as having a local stress tensor, in a way similar with particle having spins. This representation is powerful enough to describe the liquid phase in the high temperature regime \cite{ES82}. In this view, a liquid appears as a free gas of objects represented by non interacting atomic stress tensors. Numerical simulations \cite{ES82} suggest that this atomic stress tensor is a random Gaussian variable with zero average and with a covariance matrix given by a Maxwell-Bolztman factor using the expression of the elastic energy. It is even quantitatively better to use an effective elastic energy by taking into account the long distance effect of stress propagation using the mean field theory developed by Eshelby \cite{Esh57}, in analogy with the Clausius-Mossoty theory of effective electric permittivity in dielectric. The equilibrium state of such a free gas can then be explicitly computed using Statistical Mechanics leading to a law of Dulong and Petit for the heat capacity, an observation made for a long time, which found an explanation with this theory.

 \subsection{A Toy Model}
 \label{visc.ssect-toy}

 What happens at lower temperature~? The atomic vibrations, leading to acoustic waves, which will be called abusively {\em phonons} here, whereas quantized or classical, are still existing in liquids as in solids. However the acoustic waves with short wave length, namely comparable to the inter-atomic distance, are strongly damped in the liquid phase at high temperature. Whenever the typical time scale characterizing the jump dynamics is comparable to the phonon period, the phonon has no time to oscillate and it is damped. The present model investigates this damping more quantitatively. To proceed, the same intuition will be followed as in the original paper of Drude on electronic transport in metals \cite{Dr1900}. In order to make the model computable, it will be assumed that it describes the mechanical fate of a given atom: (i) as long as no jump occurs, this atom is located near the minimum of a local potential well, and it will oscillate, harmonically, with a frequency given by the local curvature of the well, (ii) then at random times (Poissonian distribution) the atom jumps, finding itself, after the jump in a new potential well, with a new local curvature and a new initial  location relative to the new equilibrium position and a new initial velocity (phase-space position). Hence, the curvature of potential well (expressed in terms of the frequency of oscillation) will be considered as a random variable with a fixed average and a given covariance $\sigma$ which will be shown to be the most important parameter. Similarly, the new relative initial phase-space position, after the jump, will also be chosen randomly according to a Maxwell-Boltzmann distribution. To make the model even more computable, it will be assumed that only one degree of freedom of oscillation is allowed. It will be seen that, even though this model is so simplistic, it exhibits a transition between two time-scales as $\sigma$ decreases. The computation of correlation functions, like the viscosity, shows that this change of scale may affect the viscosity in an essential way.

 \subsection{Anankeons in Glass: the STZ-Theory}
 \label{visc.ssect-stz}

 For a very long times, engineers investigating the mechanical properties of materials measured the strain as the material is submitted to an external stress. The strain-stress curves shows usually a line at small stress, corresponding to the elastic response. This response is usually reversible: increasing then decreasing the stress follows the same curve. If the stress applied becomes large most materials just break. But some materials, like bulk metallic glasses, exhibit a plastic region in which the sample can be strained irreversibly, without much of additional stress. In a series of seminal papers \cite{Fa98,FL98,LaPe03,Lan04,Lan06}, Langer and his collaborators developed a theory of what they called {\em shear transformation zone} (STZ), to account for the plasticity region in bulk metallic glasses. The idea is that in the plastic region, the local clusters exhibit droplets in which the solid really behaves like a liquid, called an STZ. Considering the size of such droplets as a random variable, acting as a local lubricant, Langer {\em et al.} proposed a system of effective equation at the macroscopic scale that are modeling the behavior of the material under stress with an amazing accuracy \cite{RB12}. At the atomic scale though, numerical simulation show that these STZ's are the result of a cascade or avalanche of atomic swaps, namely of anankeons, \cite{DA05}, starting at a random site, called the germ. This initial anankeon induces other in its neighborhood, because of the disruption created by the propagation of the stress induced by such swaps. The cascade looks like a local catastrophic event, occurring during a very short time, leading to the building of the local liquid-like droplets. The time scale representing the time necessary to create an STZ is however long in comparison with the typical time scale of the anankeon-swap. A model for such a cascade of event was proposed in \cite{SDM01,FIE15}.

\section{Liquids at Large Temperature as a Perfect Anankeon Gas}
\label{visc.sect-perfect}

 The numerical simulation produced by Egami and Srolovitz \cite{ES82} can be interpreted as a description of the liquid phase, far from the liquid-solid transition, in terms of a free gas of anankeon. The degree of freedom associated with each anankeon will be given in terms of an atomic stress tensor. Their Gaussian distribution observed in the numerical results can be interpreted in terms of a Gibbs state describing the thermal equilibrium in a statistical mechanical approach. This Section is dedicated to describe this in detail.

 \subsection{The Delone Hypothesis and the Ergodic Paradox}
 \label{visc.ssect-DelHyp}

 An instantaneous snapshot of the atomic arrangement can be described as a set of points $\Ll$, representing the position of atomic nuclei. With a very good approximation $\Ll$ is efficiently described as a {\em Delone set}. Namely it is characterized by two length scales $0<r\leq R<\infty$ defined as follows: (i) any open Euclidean ball of radius $r$ contains at most one point of $\Ll$, (ii) any closed Euclidean ball of radius $R$ contains at least one point of $\Ll$. The first condition is equivalent to having a minimum distance $2r$ between any pair of distinct atoms. The second condition is equivalent to forbidding holes a diameter larger than $2R$. 

\vspace{.1cm}

 However, the principles of Equilibrium Statistical Mechanics and the ergodicity with respect to translations contradict such an hypothesis.  This is the content of the so-called {\em Ergodicity Paradox} \cite{Bel15}. This is because in the infinite volume limit, the ergodic theorem implies that with probability one, given any $\epsilon >0$, somewhere in the atomic configuration two atoms are at a distance shorter than $\epsilon$. Similarly, with probability one, given any $\epsilon >0$, somewhere in the atomic configuration there is a hole of size larger than $1/\epsilon$. However, such local arrangement are rare, so that dynamically their lifetime is so short that they are unobservable. Hence restricting the atomic configurations to make up a Delone set is a reasonable assumption called the {\em Delone Hypothesis}

 \subsection{Voronoi Tiling, Delone Graphs and Local Topology}
 \label{visc.ssect-VDLT}

 Given a Delone set $\Ll$ and a point $x\in\Ll$, the Voronoi cell $V(x)$ is defined as the set of points in space closer to $x$ than to any other point in $\Ll$. $V(x)$ can be shown to be the interior of a convex polyhedron containing the open ball centered at $x$ of radius $r$ and contained in the ball centered at $x$ of radius $R$ \cite{AK00,AKL13}. The closure $T(x)$ of the Voronoi cell is called a Voronoi tile. Two such tiles can only intersect along a common face. In particular two Voronoi cells centered at two distinct points of $\Ll$ do not intersect. The entire space is covered by such tiles, so that the family of Voronoi tiles constitute a tiling, the {\em Voronoi tiling}. The vertices of the Voronoi tiles will be called {\em Voronoi points}. Their set $\Ll^\ast$ is usual called the {\em dual lattice}. For a generic atomic configuration, a Voronoi point has not more than $d+1$ atomic neighbors in space-dimension $d$ ($3$ in the plane and $4$ in the space). In particular, the Voronoi points are the center of a ball with $d+1$ atoms on its boundary, a property called by Delone (who signed Delaunay) the {\em ``empty sphere property''} \cite{Del34}. It means that it is at the intersection of $d+1$ Voronoi tiles. The $d+1$ atoms located at the center of those tiles are the vertices of a simplex (triangle for $d=2$, tetrahedron for $d=3$) that generates the so-called {\em Delaunay  triangulation}.

\vspace{.1cm}

 Two points $x,y\in\Ll$ will be called nearest neighbors whenever their respective Voronoi tiles are intersecting along a facet, namely a face of codimension 1. The pair $\{x,y\}$ will be called an {\em edge}. In practice, though, using physical criteria instead of a geometrical one, an edge might be replaced by the concept of {\em bond}, to account for the fact that the atoms associates with $x$ and $y$ are strongly bound \cite{Be59,Be60,Be64}. The term ``strongly'' is defined through  a convention about the binding energy, so as to neglect bonds between atoms weakly bound to each other. Whatever the definition of an edge, geometrical or physical (bonds), this gives the Delone graph $\gs=(\Ll, \es)$, where $\Ll$ plays the role of the vertices and $\es$ denotes the set of edges or of bonds. At this point the Delone graph needs not being oriented. 

\vspace{.1cm}

 Along a graph, a path is an ordered finite set of edges, each sharing a vertex with its successor or with its predecessor. The number of such edges is the length of the path. The graph-distance between two vertices is the length of the shortest paths joining them. A {\em graph-ball} centered at $x$ of radius $n$ will be the set of all vertices at graph distance at most $n$ from $x$. Such graph balls correspond to local clusters. Two graph balls are called {\em isomorphic} if there is a one-to-one surjective map between their set of vertices, that preserve the graph distance. As a result, if the atomic positions is slightly changed locally in space, the graph balls might still be isomorphic. Therefore graph ball, modulo isomorphism are encoding what physicists called the {\em local topology} of the atomic configuration. 

\vspace{.1cm}

 The transition between two graph balls can be described precisely through the concept of {\em Pachner moves} \cite{Pa91,AK00,AKL13}. This concept was created by experts of computational geometry, to describe the deformation of manifolds via a computer. A Pachner moves can only occur if at least one Voronoi point becomes degenerate, namely it admits at least $d+2$ atomic neighbor at some point during the move. Such moves correspond to a local change in a graph ball, namely an edge disappear another may reappear. A Pachner move corresponds exactly to the concept of anankeon. It can be described through using the Delaunay triangulation as explained in Fig.~\ref{visc.fig-pachner}.

\begin{minipage}[b]{7.5cm}
\begin{center}
\includegraphics[width=7cm]{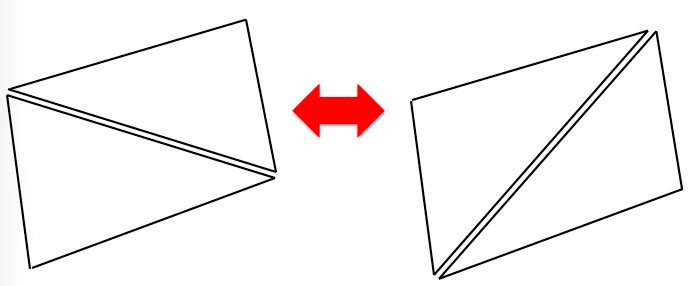}
\end{center}
\end{minipage}
\begin{minipage}[b]{7.5cm}
\begin{center}
\includegraphics[width=7cm]{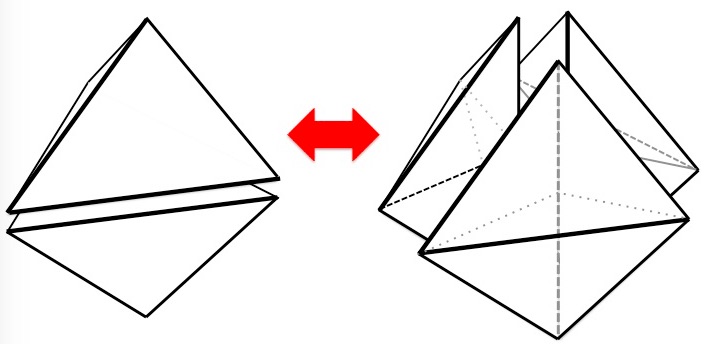}
\end{center}
\end{minipage}
\begin{figure}[ht]
   \centering
\caption{\label{visc.fig-pachner} Left: a Pachner move in the plane. 
Right: a Pachner move in the $3$-space}
\end{figure}

\begin{figure}[ht]
   \centering
\includegraphics[width=9cm]{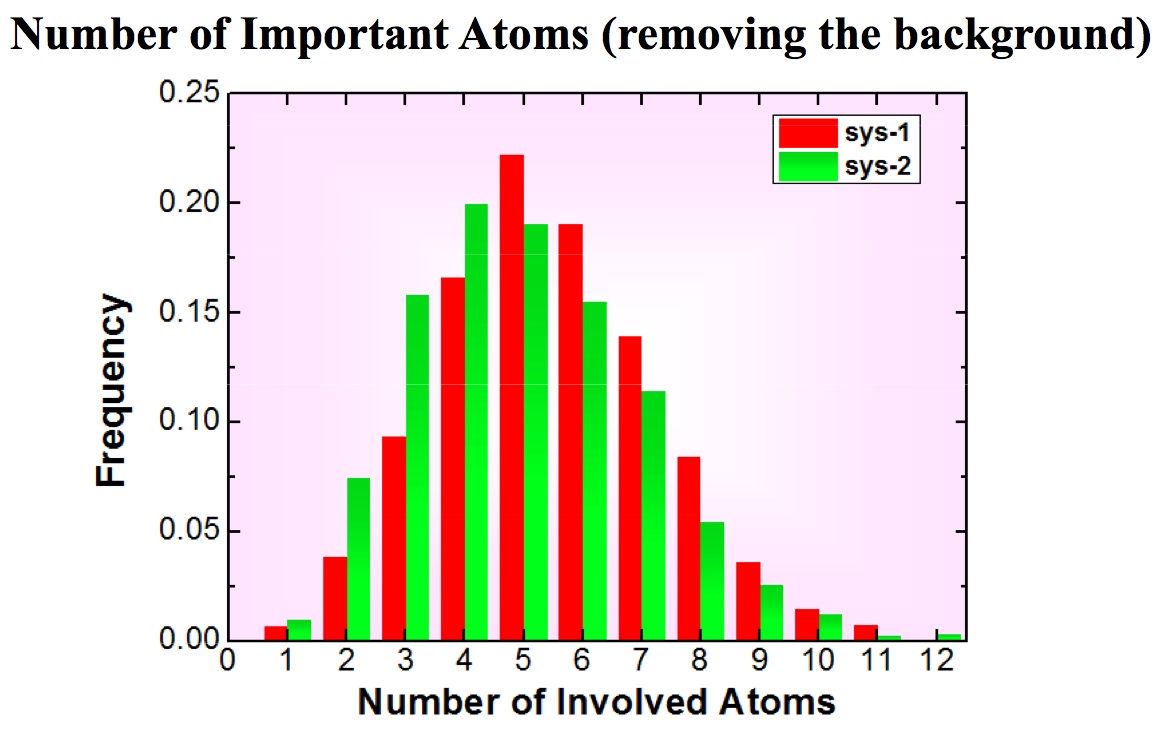}
\caption{\label{visc.fig-yue} Number of atoms involved in a Pachner move \cite{Yue14}}
\end{figure}

 In $3D$-space such a move involves at least five atoms moving quickly to deform the Delaunay triangulation, changing the nature of the Delone graph, namely the local topology. This number five has been seen in numerical simulations \cite{Yue14} as can be seen in Fig.~\ref{visc.fig-yue}.

 \subsection{Atomic Stress Tensor}
 \label{visc.ssect-atomstress}

 The definition of stress tensor in solids \cite{BH54,LPKL}, using the hypothesis of Continuum Mechanics, gives an intuition upon how to build a stress tensor at the atomic level \cite{EMV80}. The present presentation will use a different route, but leads to a similar result. It will be assumed that interatomic forces are described by a two-body isotropic potential $W(r)$, where $r$ denotes the distance between the two atoms. In such a case, given a pair $e=(x,y)$ of atoms bound by an edge, let $n_e$ denotes the unit vector along the line generated by $x,y$ oriented from $x$ to $y$. By isotropy, the force acting upon the atom located at $x$ coming from $y$ is $F_y(x)= W'(r) n_e$. The total force acting on $x$ is therefore given by the sum $F(x)=\sum_{y} n_e W'(r)$, where the sum includes the nearest neighbors of the atom at $x$. In Continuum Mechanics, the stress tensor is defined so as to satisfy  $F=\int_{\partial \Lambda}\sigma(s)\cdot n(s)ds$ for any volume $\Lambda$ with boundary $\partial \Lambda$. At the atomic level though, the integral will be replaced by the previous sum. The role of the volume $\Lambda$ is played by the Voronoi tile $T(x)$. Each nearest neighbor of $x$ corresponds to a unique facet of $T(x)$, and then the unit vector $n_e$ is nothing but the normalized unit vector perpendicular to the facet oriented outwardly {\em w.r.t.} the tile. This suggests the following formula for the stress tensor: each bond $e$ gives it a value $\sigma_e$ so that

$$\sigma_e=|n_e\rangle\langle n_e| \;\frac{W'(r_e)}{A_e}\,,
$$

 where $r_e$ is the Euclidean length of the bond and $A_e$ is the Euclidean area of the facet associated with $e$. It ought to be remarked that the ratio $\tr(\sigma_e)=p_e= W'(r)/A_e$ represents the pressure felt by the atom coming from its neighbor in the direction of the edge $e$. The deviatory stress is the traceless part $\sigma_e-p_e/d$ where $d$ is the dimension of the space in which this material lies. The atomic stress at $x$ is just the sum of these contributions, namely for an atom located at the vertex $x\in\Ll$, 

\begin{equation}
\label{visc.eq-atomstress}
\sigma(x)= \sum_{\partial_0e=x} 
 |n_e\rangle\langle n_e| \;\frac{W'(r_e)}{A_e}\,.
\end{equation}

 In this expression, $\partial_0e=x$ means the edges with origin at $x$. The formula above is very similar to the one obtained in \cite{EMV80}, using an argument from Continuum Mechanics, apart from a possible numerical constant close to $1$.

\vspace{.1cm}

 The numerical simulation in \cite{ES82} suggests that in the liquid phase, far from the liquid-solid transition, the statistical distribution of the atomic stress tensor follows a Maxwell-Boltzmann-Gibbs law of the form

\begin{equation}
\label{visc.eq-ES}
\Pro\{\sigma\in \hLam\}=\frac{1}{Z_1}
   \int_{\hLam} e^{-\beta(p^2/2B+\tau^2/2G)}\; d\sigma\,.
\hspace{2cm}
  \mbox{\rm\bf (Egami-Srolovitz Principle)}
\end{equation}

 In this expression, $\beta=1/\kB T$, $\hLam$ represents a volume in the space of stress tensors, $Z_1$ is a normalization factor, $p$ is the pressure given by the normalized trace of the stress tensor, and $\tau$ is called the {\em von Mises stress}. The latter is defined such that $\tau^2$ is the trace of the square of the deviatory stress (traceless part of $\sigma$). The coefficients $B,G$ are respectively the {\em bulk modulus} and the {\em shear modulus}. The expression in the exponential represents the elastic energy associated with the atomic stress.

\begin{figure}[ht]
   \centering
\includegraphics[width=11cm]{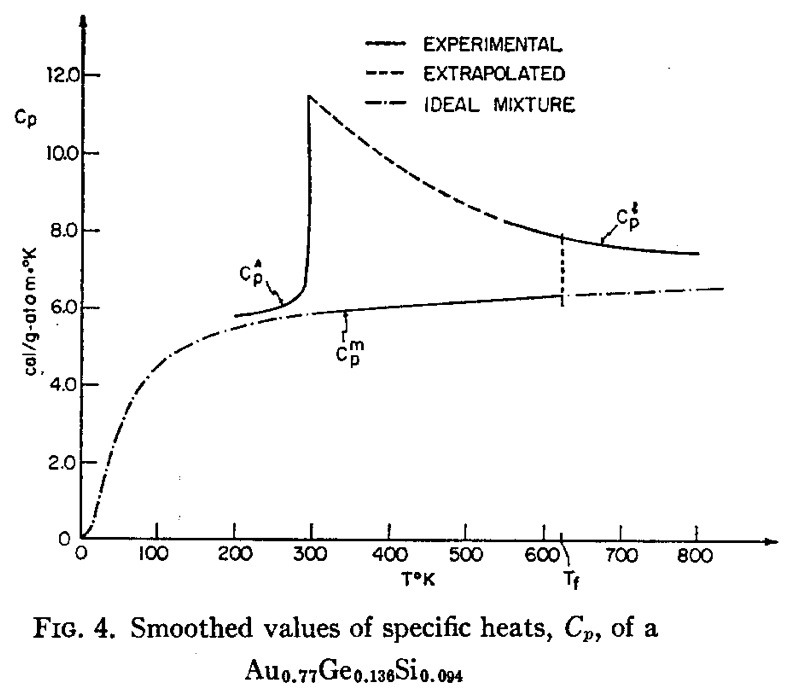}
\caption{\label{visc.fig-heatCap} Specific heat as a function of temperature for an alloy made of Gold, Germanium and Silicon signaling a glass-liquid transition \cite{CT68}}
\end{figure}

\vspace{.1cm}

 \subsection{Equilibrium for Liquid at High Temperature}
 \label{visc.ssect-liqeq}

 Under the Egami-Srolovitz principle, if the atomic stress are uncorrelated, it becomes possible to compute the partition function $Z(T,N)$ of a volume of liquid containing $N$ atoms. Namely $Z(T,N)=Z_1(T)^N$, where

$$Z_1(T)= \int e^{-\beta(p^2/2B+\tau^2/2G)}\; d\sigma\,,
   \hspace{2cm}
    \beta=\frac{1}{\kB T}\,.
$$

 The domain of integration is the space of all real symmetric $d\times d$ matrices, representing a possible stress tensor. This Gaussian integral is easy to compute giving $Z_1(T)= \zs\, T^{d(d+1)/2}$ where $\zs$ is a constant depending on $B,G$, but independent of the temperature. The Clausius entropy is given by $S=\kB \ln(Z(N,T))$ leading to the following expression of the heat capacity $C_v= c_d N\kB$ where $c_d=d(d+1)/2$, which is $6$ in $3D$. In particular, the system follows a {\em Law of Dulong-Petit} like the contribution of phonons in crystal. However, the physical origin is different in liquid as the main degree of freedom associated with atoms are the atomic stress. Hence a liquid can be seen as a {\em perfect gas of anankeons}. This prediction gave a satisfactory answer to the experimental observation \cite{CT68} of the saturation of the heat capacity at high temperature (see Fig.~\ref{visc.fig-heatCap}).

\section{Phonon-Anankeon Interaction}
\label{visc.sect-phan}

 What happens at lower temperature, near the solid-liquid transition~? The following model will show that as the temperature decreases, the vibration modes are constraining the shape of local clusters, so as to change the viscosity in a non trivial way.

 \subsection{Time Scales}
 \label{visc.ssect-tsc}

 Since the dynamics plays such an important role in explaining the difference between solids and liquids, the various relevant time scales ought to be discussed. The most elementary is $\tau_{\mbox{\em \tiny LC}}$, where $LC$ stands for {\em local configuration}, representing the time required between two consecutive Pachner moves. It corresponds to  the anankeon relaxation time. The next time scale is called {\em Maxwell's relation time} $\tau_{\mbox{\em \tiny M}}$ \cite{HMcD06} is defined by the ratio

\begin{equation}
\label{visc.eq-maxwell}
\tau_{\mbox{\em \tiny M}}=\frac{\eta}{G_\infty}\,,
\end{equation}

 where $\eta$ is the viscosity and $G_\infty$ is the high frequency shear modulus. The Maxwell relaxation time represents the time scale below which the system behaves like a solid and beyond which it can be considered as a liquid. At high temperature $\tau_M=\tlc$. Besides these two time scales, there is the $\alpha$-relaxation time $\tau_\alpha$ \cite{FKST}, defined by the scattering function at the peak of the structure function $S(Q)$. Finally there is the {\em bond lifetime} $\tau_B$ \cite{YO98}. The value of these time scales has been discussed thoroughly in \cite{IE12,INE13} (see Fig.~\ref{visc.fig-relTime}). This was done through classical as well as {\em ab initio} molecular dynamics simulation on various metallic liquids like liquid iron, $Cu_{56}Zr_{44}$ or  $Zr_{50}Cu_{40}Al_{10}$. 

\begin{minipage}[b]{7.5cm}
\begin{center}
\includegraphics[width=7cm]{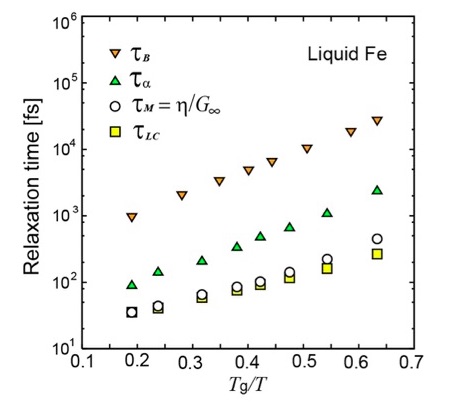}
\end{center}
\end{minipage}
\begin{minipage}[b]{7.5cm}
\begin{center}
\includegraphics[width=7cm]{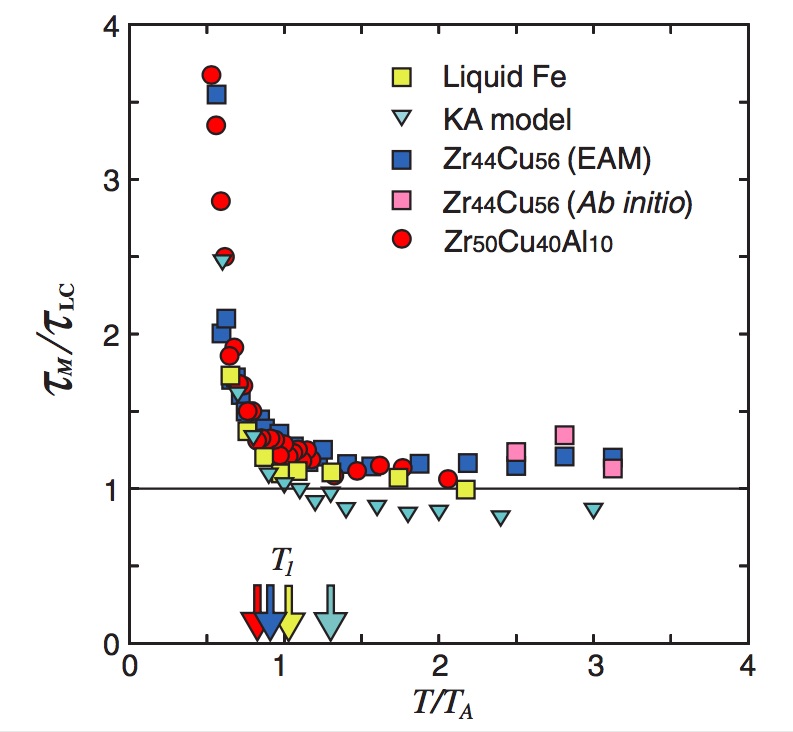}
\end{center}
\end{minipage}
\begin{figure}[ht]
   \centering
\caption{\label{visc.fig-relTime} Various relaxation times in the liquid phase as function of the temperature. Left: $T_g$ is the temperature at the liquid-solid transition. The Arrhenius law is visible on this graph: the slope of the lines give an estimate of $W$. Right: $T_A$ is the crossover temperature below which the Maxwell time differs from the $\tau_{\mbox{\em \tiny LC}}$. (\cite{INE13})}
\end{figure}

 All these time scales are proportional in the liquid phase, far enough from the liquid-solid transition as shown in Fig.~\ref{visc.fig-relTime} and are given by an Arrhenius law $\tau \sim e^{W/\kB T}$.

 However, near the liquid-glass transition point, there is a significant discrepancy between the Maxwell time and the local configuration time as can be seen on the right of  Fig.~\ref{visc.fig-relTime}. This is precisely the discrepancy that the present model is going to describe. It seems natural, then, to choose the following basic time scale for the purpose of the modeling:

\begin{itemize}
  \item\label{visc.Time-scale} {\bf Time-scale assumption: } In the toy model $\tau=\tau_{\mbox{\em \tiny LC}}$.
\end{itemize}

 \subsection{Phonon Dynamic: the main approximations}
 \label{visc.ssect-phdynapp}

 As explained in Section~\ref{visc.ssect-toy}, the model will describe the situation of one atom in the liquid. Most of the time, this atom is located near the bottom of a potential well, created by the interaction with other atoms around. Since the potential is smooth near its minimum, there is no loss of generality in assuming that it is well approximated by a local paraboloid, at least if the atom does not move too far from the minimum. Hence the atomic motion is well approximated by an harmonic oscillator. To simplify further, it will be assumed that {\em only one of the three dimension really matter}. This looks drastic, but the method of solving the model is actually the same if all the dimensions are taken into account. Hence, the curvature of the well near the minimum defines the oscillator pulsation $\omega=2\pi\nu$ where $\nu$ is the frequency. Let then $q=q(t)$ denotes the $1D$ distance of the atom from the potential minimum at any given time $t$. It is convenient to introduce the variable $u=\omega q$, which has the same dimension as the velocity $v=\stackrel{\cdot}{q}$. Hence the $2D$-vector $X$ with coordinate $u,v$ represents the phase-space position of the atom at any given time. The usual equation of motion is given by 

$$m\frac{d^2q}{dt^2}+ kq^2=0\,,
   \hspace{2cm}
    \omega= \sqrt{\frac{k}{m}}\,,
$$

  where $m$ is the mass of the atom and $k$ is the spring constant, expressible in terms of the second derivative of the potential energy. This equation can equivalently be written as

\begin{equation}
\label{visc.eq-harmosc}
\frac{dX}{dt}=\omega J X\,,
  \hspace{2cm}
   J=\left[
\begin{array}{cc}
~0 & 1\\
-1 & 0
\end{array}
\right]\,.
\end{equation}

 The solution of this equation, with initial position $u(0)$ and initial velocity $v(0)$, is given by

\begin{equation}
\label{visc.eq-solharm}
X(t)=e^{\omega t J} X(0)= 
\left[
\begin{array}{cc}
\cos{\omega t} & \sin{\omega t}\\
-\sin{\omega t} & \cos{\omega t}
\end{array}
\right]\;\left[
\begin{array}{c}
u(0)\\
v(0)
\end{array}
\right]\,.
\end{equation}

 The mechanical energy of this oscillator is given by 

$$E= \frac{m\stackrel{\cdot}{q}^2}{2}+\frac{k q^2}{2}=
   \frac{m|X|^2}{2}\,.
$$

 It ought to be stressed that the spring constant $k$ depends in an essential way upon the local environment seen by the test-atom. In particular, if the local configuration changes due to a Pachner move, or an anankeon, this constant will change as well. This will be described mathematically in the next paragraph.

 \subsection{The toy model for anankeon-phonon interaction}
 \label{visc.ssect-anandyn}

 When an anankeon strikes, the atom is ejected from its position to jump on a position nearby, in a new harmonic potential. To describe such a situation, following the strategy described in Section~\ref{visc.ssect-toy}, it will be assumed that the anankeons strikes at {\em random times} $\{ \cdots <\tau_n <\tau_{n+1}<\cdots\}$. Here the integer index $n$ varies from $-\infty$ to $+\infty$. To be more precise, the randomness of those times will be assumed to be {\em Poissonian}, namely the random variables $\tau_{n+1}-\tau_n$ are {\em i.i.d.}, with average

$$\EM\{\tau_{n+1}-\tau_n\}\stackrel{def}{=}
   \langle \tau_{n+1}-\tau_n\rangle= \tau =\tlc\,,
$$

 and an exponential distribution

$$\Pro\{(\tau_{n+1}-\tau_n)\in A\subset \RM\}= 
   \int_A e^{-t/\tau} \; \frac{dt}{\tau}\,.
$$

 The phase-space position $X(t)$ after the jump will also be relative to the minimum of the new potential well. In particular, there is a need to update this position right after the jump in terms of the position right before, namely

$$X(\tau_n+0)=X(\tau_n-0)+\xi_n\,.
$$

 The vector $\xi_n$ represents the result of the anankeon kick 
on the initial phase-space position, relative to the potential minimum. It is reasonable to assume that $\xi_n$ is also a random variable and that the family $(\xi_n)_{n\in\ZM}$ is made of {\em i.i.d.}'s. In addition it will be assumed that the common distribution is provided by the Gibbs state at the temperature $T$. Hence

\begin{equation}
\label{visc.eq-gibbs}
\Pro\{\xi_n \in B\subset \RM^2\}= 
   \frac{\kB T}{2\pi m}
    \int_B e^{-m|\xi|^2/\kB T} \; d^2\xi\,.
\end{equation}

 At last, the pulsation $\omega$ will depend upon the potential well in which the atom fell. Hence between the times $\tau_n$ and $\tau_{n+1}$ it will be given by $\omega_n$. Since the atom this model is describing should be typical, the $\omega_n$'s should be considered as random variables, and again, the family $(\omega_n)_{n\in\ZM}$ will be made of {\em i.i.d.}'s. At this point, it is unwise to fix the common distribution. It enough to fix the average $\omega$ and the variance $\sigma$

\begin{equation}
\label{visc.eq-omega}
\omega = \EM\{\omega_n\}\,,
 \hspace{2cm}
  \sigma=\Var(\omega_n)= \EM\{(\omega_n-\omega)^2\}^{1/2}\,.
\end{equation}

 At last, during the time interval $(\tau_n,\tau_{n+1})$, the evolution of $X(t)$ will satisfy eq.~(\ref{visc.eq-harmosc}) and eq.~(\ref{visc.eq-solharm}).

  \subsection{Viscosity and correlation function}
  \label{visc.ssect-visc}

 The viscosity is given by the fluctuation-dissipation theorem as

\begin{equation}
\label{visc.eq-visc}
\eta = \frac{V}{\kB T}
   \int_0^\infty C_f(t)\, dt\,,
\end{equation}

 where $V$ is the volume of the fluid, $T$ the temperature, and $C_f$ is a correlation functions defined by

\begin{equation}
\label{visc.eq-corr2}
C_f(t) =
   \EM\left\{ f(X(t))\;\overline{f(X(0)} \right\}\,,
\end{equation}

 where $f$ is a function on the  phase space giving the expression of the stress tensor. For the purpose of this calculation, it is sufficient to consider $f$ as a scalar valued function, because it does not change the conclusion of the computation. The goal will be to show that, as $t\to\infty$, the correlation $C_f(t)\sim e^{-t/\tau_M}$ defining the Maxwell time $\tau_M$. And indeed the time integral in eq.~(\ref{visc.eq-visc}) will be proportional to $\tau_M$ as anticipated by eq.~(\ref{visc.eq-maxwell}).

\vspace{.1cm}

 This correlation function can also be computed using the {\em dissipative evolution} defined by the operator $P_t$ as the following conditional expectation

\begin{equation}
\label{visc.eq-sg}
P_tf(x) = \EM\left\{ f(X(t))|X(0)=x\right\}\,.
\end{equation}

 Then, the correlation function is given by the phase-space integral

\begin{equation}
\label{visc.eq-corrHilb}
C_f(t) =\int_{\RM^2} \overline{f(x)}\,P_tf(x)\;d^2x\,.
\end{equation}

 In order to compute $\tau_M$ precisely, it is convenient to compute the Laplace transform of $C_f$ instead, namely, given a {\em complex variable} $\zeta$

\begin{equation}
\label{visc.eq-lapcorr}
\ls C_f(\zeta)=
   \int_0^\infty e^{-t\zeta}\, C_f(t)\, dt\,.
\end{equation}

 It turns out that the asymptotic behavior of the correlation at $t\to \infty$ can be interpreted as the position of the nearest singularity of $\ls C_f$ in the complex plane. Namely the Laplace transform is analytic in the half-plane $\Re\zeta >-1/\tau_M$. Hence, in order to compute the correlation function, it is sufficient to compute the value $f(X(t))$ taken by the function $f$ along the random phase-space trajectory of the atom under scrutiny.

\section{Computing the Laplace transform of a correlation function}
\label{visc.sect-corrcomp}

\noindent The formula~(\ref{visc.eq-corrHilb}) suggests to compute the Laplace transform of the dissipative evolution operator $P_t$ instead, acting on some function space describing the relevant functions on the phase-space. In order to do so, it will be more efficient to compute the actions of the various part of the dynamic on functions rather than on the phase-space trajectory. As will be seen this calculus is very similar to the operator calculus used in Quantum Mechanics.

 \subsection{Operator Calculus}
 \label{visc.ssect-opcal}

\noindent The first part of the evolution is provided by the phase-space mouvement of the harmonic oscillator, as described in Section~\ref{visc.ssect-phdynapp}. Namely

\begin{equation}
\label{visc.eq-angmom}
f\left(e^{\omega t J}x\right)=
   \left(e^{-\omega t\JM}f\right)(x)\,,
    \hspace{2cm}
     -\JM=v\partial_u-u\partial_v\,,
\end{equation}

\noindent where $x=(u,v)$ is the phase-space position and $\JM$ is very similar to the {\em phase-space angular momentum} but for the missing $\imath$ in front (namely $\JM$ is anti-Hermitian instead) and its $2D$ character in phase-space.

\vspace{.1cm}

\noindent Similarly, the impact of the anankeon on the atom is to translate the phase-space position of the atom after a kick, by a vector $\xi=(a,b)$, leading to the {\em translation operator}

\begin{equation}
\label{visc.eq-trans}
f(x+\xi)= \left(e^{\xi\cdot \nabla}f\right)(x)\,,
    \hspace{2cm}
     \xi\cdot \nabla= a\partial_u+b\partial_v\,.
\end{equation}

\noindent Hence $\nabla$ plays the role of a phase-space momentum operator apart for the missing $\imath$ in front (namely $\nabla$ is anti-Hermitian instead).

\vspace{.1cm}

\noindent Let $X(t)$ denote the phase-space trajectory of the atom at time $t$ with initial condition $X(0)=x$. From the description of the model, an harmonic motion takes place between time $\tau_0=0$ to $\tau_1$ at pulsation $\omega_1$, then a phase-space translation by $\xi_1$, then another harmonic motion between time $\tau_1$ to $\tau_2$ at pulsation $\omega_2$, then a new phase-space translation by $\xi_2$, and so on, until the time $t$ is reached. It becomes possible to describe it from the functional point of view as follows

\begin{equation}
\label{visc.eq-evol}
P_tf(x)= \EM\left(f(X(t))|X(0)=x\right) =
\EM\left\{
 \prod_{j=1}^n
  \left(e^{-(\tau_j-\tau_{j-1})\omega_{j-1}\JM}\, 
        e^{\xi_j\cdot\nabla}
  \right)
    e^{-(t-\tau_n)\omega_n\JM}\,f
        \right\}(x)\,,
\end{equation}

\noindent where the product is ordered from left to right, with $j=1$ on the left and $j=n$ on the right, and where $n$ is the integer such that $\tau_n\leq t<\tau_{n+1}$. From the assumptions made on the model, the $(\tau_j-\tau_{j-1})$ are {\em i.i.d.}, as well as the $\omega_j$'s and the $\xi_j$'s. It follows that $n$ is also a random variable. Computing the distribution of $n$ though, is hard enough to try to avoid it. The Laplace transform will allow to avoid such a calculation.

 \subsection{Laplace transformation}
 \label{visc.ssect-laping}

\noindent After averaging and taking the Laplace transform the result is

$$\ls P_\zeta f(x)=
   \int_0^\infty e^{-t\zeta}\; P_tf(x)\;dt=
    \EM\left\{\int_0^\infty e^{-t\zeta}\;f(X(t)\;dt \Bigg| X(0)=x\right\}\,,
     \hspace{1cm}
      \zeta\in \CM\,.
$$

\noindent Thanks to the linearity of the averaging process, this integral over time can be decomposed, under the averaging, into the time intervals $[\tau_n,\tau_{n+1}]$, to give

$$\ls P_\zeta f(x)=\sum_{n=0}^\infty
    \EM\left\{\int_{\tau_n}^{\tau_{n+1}} 
     e^{-t\zeta}\;f(X(t)\;dt\Bigg| X(0)=x\right\}\,.
$$

\noindent  Using eq.~(\ref{visc.eq-evol}), this gives

\begin{equation}
\label{visc.eq-ev}
\ls P_\zeta f(x)= \sum_{n=0}^\infty
    \EM\left\{\int_{\tau_n}^{\tau_{n+1}} 
     e^{-t\zeta}\; \prod_{j=1}^n
      \left(e^{-(\tau_j-\tau_{j-1})\omega_{j-1}\JM}\, 
        e^{\xi_j\cdot\nabla}
      \right)
       e^{-(t-\tau_n)\omega_n\JM}\,f(x)\;dt
        \right\}\,.
\end{equation}

\noindent Since $\tau_0=0$, a telescopic sum give $t=(t-\tau_n)+(\tau_n-\tau_{n-1})+\cdots+ (\tau_1-\tau_0)$. This leads to distribute the first term $e^{-t\zeta}$ inside each terms of the product leading to replace each $\omega_{j-1}\JM$ by $\omega_{j-1}\JM + \zeta$, namely

\begin{equation}
\label{visc.eq-ev2}
\ls P_\zeta f(x)= \sum_{n=0}^\infty
    \EM\left\{\int_{\tau_n}^{\tau_{n+1}} 
     \prod_{j=1}^n
      \left(e^{-(\tau_j-\tau_{j-1})(\omega_{j-1}\JM+\zeta)}\, 
        e^{\xi_j\cdot\nabla}
      \right)
       e^{-(t-\tau_n)(\omega_n\JM+\zeta)}\,f(x)\;dt
        \right\}\,.
\end{equation}

\noindent The first operation consists in evaluating the integral over $t$. The change of variable $s=t-\tau_n$ gives the following operator valued integral

\begin{equation}
\label{visc.eq-sint}
\int_0^{\tau_{n+1}-\tau_n}
    e^{-s(\omega_n\JM+\zeta)}\;ds= 
     \frac{1-e^{-(\tau_{n+1}-\tau_n)(\omega_n\JM+\zeta)}}
       {\zeta+ \omega_n\JM}\,.
\end{equation}

\noindent The next step consists in recognizing that the averaging $\EM$ can be decomposed into averaging over the time Poisson process, before averaging over the random phase-space positions $\xi_j$'s or the random pulsations $\omega_{j-1}$'s. In addition, the stochastic independence of the $(\tau_j-\tau_{j-1})$'s permits to reduce the averaging to each factor 
in the product in eq.~(\ref{visc.eq-ev2}). Using the the {\em time-scale assumption} in Section~\ref{visc.Time-scale} on page~\pageref{visc.Time-scale}, this averaging reduces to the following integral, where $A$ is an operator

\begin{equation}
\label{visc.eq-pint}
\EM_\tau\left\{e^{-(\tau_j-\tau_{j-1})A}\right\}=
   \int_0^\infty e^{-s/\tau-sA}\;\frac{ds}{\tau} =
    \frac{1}{1+\tau A}\,,
\end{equation}

\noindent where $\EM_\tau$ denotes the average over the Poissonian times. The stochastic independence of each factor in the product, leads to a factorization of the averaging, since, whenever two random variable $X,Y$ are stochastically independent the average of their product is given by the product of their average, that is $\EM(XY)=\EM(X)\EM(Y)$. After a little algebra, using eq.~(\ref{visc.eq-sint}, \ref{visc.eq-pint}), the eq.~(\ref{visc.eq-ev2}) leads to

\begin{equation}
\label{visc.eq-ev3}
\ls P_\zeta f(x)= \tau \sum_{n=0}^\infty
    \EM\left\{ 
     \prod_{j=1}^n
      \left(
        \frac{1}{1+\tau(\zeta+\omega_{j-1}\JM)}\, 
        e^{\xi_j\cdot\nabla}
      \right)
       \frac{1}{1+\tau(\zeta+\omega_n\JM)}\,f(x)
        \right\}\,.
\end{equation}

\noindent Similarly, since the $\xi_j$'s and the $\omega_{j-1}$'s are stochastically independent, it is enough to compute the following averages

\begin{equation}
\label{visc.eq-xiaver}
\EM_\xi\left\{e^{\xi_j\cdot\nabla}\right\}=
  e^{\kB T\,\Delta/2m}\,,
   \hspace{2cm}
    \Delta=\nabla\cdot\nabla= \partial_u^2+\partial_v^2\,,
\end{equation}

\noindent where $\EM_\xi$ denotes the average over the phase-space position, using the Gaussian integral in eq.~(\ref{visc.eq-gibbs}). Similarly, the average over the $\omega_{j-1}$'s leads to the operator-valued function

\begin{equation}
\label{visc.eq-omaver}
\AM(\zeta)=
 \EM_\omega\left\{
  \frac{1}{1+\tau(\zeta+\omega_{j-1}\JM)}
            \right\}\,.
\end{equation}

\noindent Evaluating this integral requires to know the distribution of the $\omega_{j-1}$'s. At this point, no further assumption will be made yet. Using eq.~(\ref{visc.eq-xiaver}, \ref{visc.eq-omaver}), the eq.~(\ref{visc.eq-ev3}) leads to

\begin{equation}
\label{visc.eq-ev4}
\ls P_\zeta f(x)= \tau \sum_{n=0}^\infty
    \left\{ 
     \AM(\zeta)\,e^{\kB T\,\Delta/2m}
    \right\}^n
      \AM(\zeta)f(x)=\tau\,
       \frac{1}{1-\AM(\zeta)\,e^{\kB T\,\Delta/2m}}\,\AM(\zeta)f(x)\,.
\end{equation}

\noindent The last series converges only if $\|\AM(\zeta)\,e^{\kB T\,\Delta/2m}\|<1$ where $\|\cdot\|$ denotes the operator norm. At this point though, no accurate description of the space on which these operators act has been done, so that the norm condition is still meaningless. It will be the purpose of the discussion of the next Section. With this cautionary statement in mind, the problem is reduced to analyzing the properties of one operator at the end, giving the Laplace transform as a function of $\zeta$.

 \subsection{Digression: choosing the function space}
 \label{visc.ssect-fspace}

\noindent In order to make sense of the calculation made in the previous Section~\ref{visc.ssect-laping}, as was remarked at the end, the function space on which these operators act must be made more precise. In order to do so, the first question is to know which type of function $f$ are physically relevant. In the calculation of viscosity, the ideal function should be the atomic deviatory stress tensor as given in eq.~(\ref{visc.eq-atomstress}). Using the finite sum decomposition over the edges linking the atom at $x$ with its neighbors, it boils down to find a space containing the functions of the form

\begin{equation}
\label{visc.eq-function}
g(x) = \frac{|y-x\rangle\langle y-x|}{|y-x|^2}\, W'(|y-x|)\,,
\end{equation}

\noindent where $y$ is a fixed point representing the position of a neighboring atom. First this expression is matrix valued instead of being a scalar. This matrix is $d\times d$ if the liquid is lying in a $d$-dimensional space. It is real symmetric by construction. Moreover, {\em only the traceless part} of this matrix matters for the viscosity. In $d$-dimension the space of $d\times d$ real symmetric matrices is invariant by the rotation group $SO(d)$, and it decomposes into the direct sum of two invariant subspaces, one corresponding to the trace (trivial representation of $SO(d)$) and the traceless part (unit representation of $SO(d)$). For $d=3$ it is well known that the unit representation of $SO(3)$, acting on this space of traceless real symmetric matrices, corresponds to the spin $2$ representation of $SU(2)$.

\vspace{.1cm}

\noindent On the other hand, the effective potential energy $W$ decays fast enough at infinity. The decay is mostly exponential due to screening effect of negatively charges electrons over the Coulomb potential of the positively charged nucleus. Moreover, it derivative is likely to share the same property. Hence, the function $g$ decays fast enough as the position $x$ of the atom diverges from its equilibrium position.

\vspace{.1cm}

\noindent The previous considerations motivated by the physical problem, have to be translated into the langage of the phase-space for the $1D$ harmonic oscillator replacing the atomic motion. In the previous Sections, the letter $x$ has been confusingly used also to describe the phase-space position $x=(u,v)$, where $u$ represents the $1D$ position of the harmonic oscillator relative to the origin chosen at the minimum of the potential well, while $v$ describes the corresponding velocity. The translation between the language used in $d$ dimensions and the phase-space language will be done in the following way:

\vspace{.1cm}

(i) The space $\Hh$ of relevant functions will be chosen so that any of them decay fast enough at infinity;

(ii) the space $\Hh$ needs to be mathematically convenient, in particular, choosing it to be a Hilbert space is a very convenient choice;

(iii) the symmetry properties of the function $f$ used in the calculation of the viscosity must reflect as much as possible the ones of the realistic stress tensor given in eq.~(\ref{visc.eq-atomstress}).

\vspace{.1cm}

\noindent A reasonable choice is $\Hh=L^2(\RM^2)$, which denotes the space of complex valued square integrable measurable functions over the phase-space $\RM^2$, namely which satisfy 

$$\|f\|_\Hh^2= \int_{\RM^2} |f(x)|^2\,d^2x <\infty\,.
$$

\noindent The symmetry properties might be reflected by the presence of the operator $\JM$, generating the phase-space rotations, namely the group $U(1)$ of rotations in the plane. It is well known that $\JM$ can be defined as an anti-selfadjoint operator on $\Hh$ with eigenvalues $\imath \ell$ where $\ell \in\ZM$ takes on any nonnegative of nonpositive integer values. Moreover the eigenvectors are given by functions of the form (using polar coordinates)

\begin{equation}
\label{visc.eq-evell}
g_\ell(x) =e^{\imath \ell \theta} g(r)\,,
   \hspace{2cm}
    x=(r\cos{\ell\theta},r\sin{\ell\theta})\in\RM^2\,,
\end{equation}

\noindent where

\begin{equation}
\label{visc.eq-L2int}
\int_0^\infty |g(r)|^2\,rdr <\infty\,.
\end{equation}

\noindent In particular this eigenspace is infinite dimensional. Let $\Pi_\ell$ denote the orthogonal projection from $\Hh$ onto this space. Moreover, these spaces are mutually orthogonal and then span the entire Hilbert space $\Hh$. 

\vspace{.1cm}

\noindent The analogy between the realistic situation and the phase-space modeling, suggests to choose the eigenspace of $\JM$ with angular momenta $\ell=\pm 2$. It will be seen though, that this restriction does not matter in terms of the qualitative features of the model.

\vspace{.1cm}

\noindent Similarly the operator $\Delta$ becomes self-adjoint with nonpositive generalized eigenvalues. In particular, the operator $\exp\{\kB T \Delta/2m\}$ is selfadjoint with its spectrum given by the interval $[0,1]$, so that $0\leq \exp\{\kB T \Delta/2m\}\leq 1$ implying $\|\exp\{\kB T \Delta/2m\}\|\leq 1$.

\vspace{.1cm}

\noindent The previous discussion shows that, while a choice has to be made for $\Hh$, it will always be partly arbitrary. In particular, the discussion of symmetries might not be totally realistic. But the choice of $\Hh$ is not innocent. The properties of the unbounded operators $\JM$ and $\Delta$ depend crucially upon which choice is made. Hence, the main criterion is whether or not the model is effective at describing the physical properties of the system under study.

 \subsection{Analyticity Domain}
 \label{visc.ssect-holom}

\noindent Using the assumptions made in Section~\ref{visc.ssect-fspace}, the operator $\{1+\tau(\zeta+ \omega_j \JM)\}^{-1}$ has a spectrum of isolated eigenvalues (each which infinite multiplicity) given by $\{1+\tau(\zeta + \imath\omega_j\ell)\}^{-1}$ with $\ell\in\ZM$. In particular, it has poles at

$$\zeta_\ell= -\left(\frac{1}{\tau} + \imath \ell \omega_j\right)\,.
$$

\noindent As the real valued random variable $\omega_j$ varies, though, the pole position describes a subset of the vertical line $\Re{\zeta}=-1/\tau$, the extend of which depends upon the support of the probability distribution of $\omega_j$. Hence after taking the average over $\omega_j$, this line becomes a cut in the $\zeta$-complex plane. It is convenient to introduce the following functions

\begin{equation}
\label{visc.eq-aell}
a_\ell(\zeta) =\EM
 \left\{
   \frac{1}{1+\tau(\zeta +\imath\omega_j\ell)}
 \right\}\stackrel{\Re(\zeta)\uparrow +\infty}{\simeq}
  \frac{1}{1+\tau\zeta}+ 
   O\left(\frac{1}{(1+\tau\zeta)^2}\right)\,,
  \hspace{1.5cm}
   \ell\in\ZM\,.
\end{equation}

\noindent In view of the definition of the Laplace transform (see eq.~(\ref{visc.eq-lapcorr})), it implies that only the part $\Re(\zeta)>-1/\tau$ is relevant for this analysis. Hence

\vspace{.1cm}

\noindent {\bf Result 1: } {\em $\AM(\zeta)$ and the $a_\ell(\zeta)$'s are complex analytic in the $\zeta$-complex plane in the domain $\Re(\zeta)>-1/\tau$.}

\vspace{.1cm}

\noindent Since $\JM$ is anti-selfadjoint, the operator $\AM(\zeta)$ is {\em normal}, namely it commutes with its adjoint. It can actually be expressed in terms of the eigenprojections $\Pi_\ell$ using its {\em spectral decomposition}

\begin{equation}
\label{visc.eq-spdecomp}
\AM(\zeta)= \sum_{\ell\in\ZM} a_\ell(\zeta)\;\Pi_\ell\,.
\end{equation}

\noindent Consequently its norm is given by the maximum of the modulus of its eigenvalues, namely, using $|\EM(X)|\leq \EM(|X|)$, 

\begin{equation}
\label{visc.eq-normA}
\|\AM(\zeta)\|=
   \sup_{\ell\in\ZM}
    \left|
     \EM_\omega\left\{
      \frac{1}{1+\tau(\Re{\zeta}+\imath (\Im{\zeta}+\omega_j\ell))}
              \right\}
    \right|\leq 
     \frac{1}{|1+\tau\Re(\zeta)|}\,.
\end{equation}

\noindent Going back to eq.~(\ref{visc.eq-ev4}), the Laplace transform of the stochastic evolution shows a singularity at values of $\zeta$ for which the spectrum of $\AM(\zeta)\exp\{\kB T \Delta/2m\}$ contains the number $1$. Since the product of two operators might not commute, the computation of the spectrum might be difficult. However, he Laplacian is always invariant by rotation in any dimension, so that

\vspace{.1cm}

\noindent {\bf Result 2: } {\em the operators $\JM$ and $\Delta$ commute.}

\vspace{.1cm}

\noindent More precisely, using the polar coordinates in $2D$ gives

$$\Delta= \frac{1}{r}\,\frac{\partial}{\partial r}
    \left(r\frac{\partial}{\partial r}\right)+ \frac{\JM^2}{r^2}=
     -\left(\hp_r^2+\frac{-\JM^2}{r^2}\right))\,,
$$

\noindent where $\hp_r$ is the analog of the (selfadjoint) {\em radial momentum}. On the eigensubspace $\Pi_\ell\Hh$, the operator $-\JM^2$ takes on the value $\ell^2$. So that the eq.~(\ref{visc.eq-ev4}) becomes

\begin{equation}
\label{visc.eq-ev5}
\ls P_\zeta = \tau\sum_{\ell\in\ZM}
 \frac{1}{1-a_\ell(\zeta) e^{-(\hp_r^2+\ell^2/r^2)}}\,
  a_\ell(\zeta) \Pi_\ell\,.
\end{equation}

\noindent Since the operator $\hp_r^2+\ell^2/r^2$ is known to admit $[0,\infty)$ as its spectrum, the {\em l.h.s.} is analytic in $\zeta$, in the subdomain of $\Re(\zeta)>-1/\tau$ made of values for which $a_\ell(\zeta)\notin [1,+\infty)$ for all the $\ell$'s that are relevant for the viscosity formula. Without knowing more about the distribution of the $\omega_n$'s it is not possible to go further.

 \subsection{An explicit case}
 \label{visc.ssect-expl}

\noindent There is a distribution of the pulsation that permits do get a closed formula for the $a_\ell$'s. Namely we assume that the pulsation is uniformly distributed on some finite interval $I$. Since $\EM(\omega_n)=\omega$ and $\Var(\omega_n)=\sigma$ (see eq.~(\ref{visc.eq-omega})), it follows that $I=[\omega-\sigma\sqrt{3}, \omega+\sigma\sqrt{3}]$. This gives

\begin{equation}
\label{visc.eq-aell1}
a_\ell(\zeta)= \frac{1}{2\imath\sqrt{3} \ell \sigma\tau}\,
 \ln\left\{
   \frac{1+\tau(\zeta+\imath\ell\omega)+\imath\sqrt{3}\ell\sigma\tau}
        {1+\tau(\zeta+\imath\ell\omega)-\imath\sqrt{3}\ell\sigma\tau}
    \right\}\,.
\end{equation}

\noindent It follows that $a_0=0$. Changing $\ell$ into $-\ell$ and $\omega$ into $-\omega$ do not change $a_\ell$. Hence there is no loss of generality in assuming that $\ell \geq 1$. As long as $\Im(\zeta)\neq -\ell \omega$, $\Im\left(a_\ell(\zeta)\right)\neq 0$ so that $1-a_\ell(\zeta)e^{-p}\neq 0$ for $p\geq 0$, so the Laplace transform of $P_t$ is well defined. To find the positions of the singularities it is therefore necessary to assume that $(\zeta+\imath\ell\omega)=\Re(\zeta)\in \RM$.  From the definition of $a_\ell$ in eq.~(\ref{visc.eq-aell}), the branch of the logarithm in eq.~(\ref{visc.eq-aell1}) is such that as $\Re(\zeta)\to +\infty$, it satisfies the asymptotics given in eq.~(\ref{visc.eq-aell}). Hence, using polar coordinates in the complex plane, this gives

$$1+\tau\Re(\zeta)+\imath\sqrt{3}\ell\sigma\tau=
   \rho e^{\imath \theta}\,,
    \hspace{2cm}
     \tan{\theta}= \frac{\sqrt{3}\ell\sigma\tau}{1+\tau\Re(\zeta)}\,.
$$

\noindent It ought to be remarked that, in the domain of analyticity, $1+\tau\Re(\zeta)>0$. With the asymptotics for $\Re(\zeta)$ large, since $\ell\geq 1$ the definition of $\theta$ implies $0<\theta<\frac{\pi}{2}$. To summarize,

\begin{equation}
\label{visc.eq-aell2}
a_\ell(\zeta)= \frac{\theta}{\sqrt{3}\ell\sigma\tau}\,,
 \hspace{2cm}
  \tan{\theta}= \frac{\sqrt{3}\ell\sigma\tau}{1+\tau\Re(\zeta)}\,,
   \hspace{2cm}
    0<\theta<\frac{\pi}{2}\,.
\end{equation}

\noindent The analyticity condition $a_\ell\neq [1,+\infty)$ imposed on the Laplace transform implies, whenever $\zeta+\imath\ell\omega=\Re(\zeta)$, that 

(i) if $\sqrt{3}\ell\sigma\tau\geq \pi/2$, then $a_\ell(\zeta)\neq [1,+\infty)$ for $\Re(\zeta)>-1/\tau$;

(ii) If $\sqrt{3}\ell\sigma\tau< \pi/2$, then in order that $a_\ell(\zeta)\neq [1,+\infty)$, it is necessary that $\theta<\sqrt{3}\ell\tau\sigma<\pi/2$; since the function $\theta\mapsto \tan{\theta}$ is increasing on $[0,\pi/2)$ then

\begin{equation}
\label{visc.eq-maxtime}
\Re(\zeta)>-\frac{1}{\tau_M}\,,
   \hspace{2cm}
    \frac{\tau_M}{\tau} =
     \left(
       1-\frac{\sqrt{3}\ell\tau\sigma}{\tan{\sqrt{3}\ell\tau\sigma}}
     \right)^{-1}
\end{equation}

\noindent In other words

\vspace{.1cm}

\noindent {\bf Result 3: } {\em The Maxwell time is given, in terms of $\tau=\tlc$ by the following formulae}

\begin{equation}
\label{visc.eq-maxtime2}
\frac{\tau_M}{\tlc} =
\left\{
\begin{array}{ll}
1 & \mbox{\rm if}\;\;\sqrt{3}\ell\tau\sigma\geq \pi/2\\
\left(
       1-\frac{\sqrt{3}\ell\tau\sigma}{\tan{\sqrt{3}\ell\tau\sigma}}
     \right)^{-1}>1 & \mbox{\rm otherwise}
\end{array}\right.
\end{equation}

\vspace{.5cm}

\section{Interpreting the Results}
\label{visc.sect-results}

\noindent From the mathematical computations made in the previous Section~\ref{visc.ssect-expl}, it follows that there are two regimes concerning the Maxwell time, both being controlled by the dimensionless parameter

$$K = \sqrt{3}\tlc\sigma\,,
$$

\noindent which is proportional to the variance of the distribution of the pulsations $\omega_n$, representing the curvature of the local potential energy, near its mechanical equilibrium, where the atom is located. In order to draw practical predictions for a liquid material, some physical hypothesis are required. In particular, {\em it is highly expected that the parameters $\tlc$ and $\sigma$ depend on the temperature of the liquid}. If so, the following conclusion can be made

\vspace{.1cm}

\noindent {\bf Result 4: } {\em Depending upon the material, there may or may not exist a cross-over temperature $T_{co}$, such that (a) for $T\geq T_{co}$, the anankeons dominate and the Maxwell time coincides with the local configuration time, while, (b) if $T<T_{co}$, the phonons restore some order, survive for a longer while, so that the Maxwell time becomes larger than the local configuration time.}

\vspace{.1cm}

\noindent In view of the results showed in Fig.~\ref{visc.fig-relTime} \cite{INE13}, it seems natural to assume that the time dependence of $\tlc$ be given by an {\em Arrhenius Law}, namely

$$\tlc=\tau_\infty e^{W/\kB T}\,,
   \hspace{1cm}\mbox{\rm where}\hspace{1cm}
    \tau_\infty= \lim_{T\uparrow \infty} \tlc(T)\,.
$$

\noindent In this expression, the energy $W$ can be interpreted as the height of the potential barrier to be passed over as the anankeon strikes. 
How is the variance $\sigma$ depends on the temperature~? So far no clues have been given on that. So we are reduced to speculate. By analogy and {\em without further justification, though}, the same hypothesis can be made on $\sigma$, 

$$\sigma =\sigma_\infty e^{-W_v/\kB T}\,,
   \hspace{1cm}\mbox{\rm where}\hspace{1cm}
    \sigma_\infty= \lim_{T\uparrow \infty} \sigma(T)\,.
$$

\noindent If so, it is legitimate to ask the following question: 

\vspace{.1cm}

\noindent {\bf Question: } {\em what is the meaning of the energy scale $W_v$~?}

\vspace{.1cm}

\noindent It is only possible to see that if $W_v$ is small, then the variance of the frequency distribution is weakly dependent of the temperature. If, on the other hand, $W_v$ is large enough, then this variance is strongly depressed at low temperature. It is important to remember that such an Arrhenius Law, while not justified by physical arguments, needs to be valid only in a temperature range of moderate size above the glass transition. In such a case, it is only a convenient way to parametrize the temperature behavior of the variance using only one parameter, the energy scale $W_v$. 

\vspace{.1cm}

\noindent Once these assumptions are made, it follows that the parameter $K$ behaves like

\begin{equation}
\label{visc.eq-KT}
K(T)= K_\infty e^{-(W_v-W)/\kB T}\,,
   \hspace{1cm}\mbox{\rm where}\hspace{1cm}
    K_\infty= \sqrt{3}\,\tau_\infty\, \sigma_\infty\,.
\end{equation}

\noindent The condition required for the existence of a cross-over temperature $T_{co}$ is that, $K(T)$ decreases as the temperature decreases and that $K(T_{co})\ell=\pi/2$, leading to

\begin{equation}
\label{visc.eq-Tco}
W_v>W\,, 
 \hspace{1cm}
  K_\infty >\frac{\pi}{2}\,,
   \hspace{1cm}\Rightarrow\hspace{1cm}
    \kB T_{co}= \frac{W_v-W}{\ln(2 K_\infty/\pi)}
\end{equation}

\noindent The first consequence of these hypothesis is that the difference between strong and fragile glassy materials \cite{Ang95} can be expressed by the conditions

\begin{equation}
\label{visc.eq-strfra}
\left\{
\begin{array}{ll}
W_v\leq W & \mbox{\rm for strong liquids}\\
W_v>W & \mbox{\rm for fragile liquids}
\end{array}\right.
\end{equation}

\noindent The second consequence is that {\em the ratio between the Maxwell and the local configuration times diverges exponentially fast below the cross-over temperature} (see eq.~(\ref{visc.eq-maxtime2}))

\begin{equation}
\label{visc.eq-ratio}
\frac{\tau_M}{\tlc}\;\;
 \stackrel{T\ll T_{co}}{\sim}\;\;
  \frac{1}{\ell^2K(T)^2} =
   \frac{e^{2(W_v-W)/\kB T}}{\ell^2 K_\infty^2}\,.
\end{equation}

\noindent Such a law would explain why the viscosity diverges so fast near the glass transition temperature.

\end{document}